\documentclass[onecolumn,usenatbib]{mn2e}

\usepackage{graphics}
\usepackage{epsfig}
\usepackage{amsmath}
\title[Two close sdb+dM binaries]
{Seismic evidence for non-synchronization in two close sdB+dM binaries from
{\it Kepler} photometry}

\author[H. Pablo et al.]
{Herbert Pablo$^{1,14}$,
Steven D. Kawaler$^{1,14}$,
M. D. Reed$^2$,
S. Bloemen$^{3,14}$,
S. Charpinet$^{4,14}$,
\newauthor
H. Hu$^{5,14}$,
J. Telting$^6$,
R. H. {\O}stensen$^3$,
A. S. Baran$^{1,2}$,
E. M. Green$^7$,
J. J. Hermes$^8$,
\newauthor
T. Barclay$^9$,
S. J. O'Toole$^{10}$,
Fergal Mullally$^{11}$,
D. W. Kurtz$^{12}$,
\newauthor
J. Christensen-Dalsgaard$^{13,14}$,
Douglas A. Caldwell$^{11}$,
Jessie L. Christiansen$^{11}$,
\newauthor
and
K. Kinemuchi$^{9}$\\
$^{1}$Department of Physics and Astronomy, Iowa State University, Ames, IA 50011 USA\\
$^{2}$Department of Physics, Astronomy and Materials Science,
 Missouri State University, 901 S. National, Springfield, MO 65897 USA\\
$^{3}$Instituut voor Sterrenkunde, KU~Leuven, Celestijnenlaan 200D, 3001 Leuven, Belgium\\
$^{4}$Laboratoire d'Astrophysique de Toulouse-Tarbes, Universit\'e de Toulouse, CNRS, 14 Av. E. Belin, 31400 Toulouse, France\\
$^{5}$Institute of Astronomy, The Observatories, Madingley
Road, Cambridge, CB3 0HA\\
$^{6}$Nordic Optical Telescope, 38700 Santa Cruze de La Palma, Spain\\
$^{7}$Steward Observatory, University of Arizona, Tucson, AZ 85721, USA\\
$^{8}$Department of Astronomy, University of Texas at Austin, Austin, TX 78712, USA\\
$^{9}$Bay Area Environmental Research Institute/NASA Ames Research Center, Moffett Field, CA 94035, USA\\
$^{10}$Anglo-Australian Observatory, P.O. Box 296, Epping, NSW 1710, Australia\\
$^{11}$SETI Institute/NASA Ames Research Center, Moffett Field, CA
94035\\
$^{12}$Jeremiah Horrocks Institute, University of Central Lancashire, Preston, PR1 2HE, United Kingdom\\
$^{13}$Aarhus University, Aarhus, Denmark\\
$^{14}$Kavli Institute for Theoretical Physics, Kohn Hall, University of California, Santa Barbara, CA 93106, USA}

\begin{document}

\pagerange{\pageref{firstpage}--\pageref{lastpage}} \pubyear{2012}

\maketitle
\label{firstpage}

\begin{abstract}
We report on extended photometry of two pulsating sdB stars in close binaries.  For both cases, we use rotational splitting of the pulsation frequencies to show that the sdB component rotates much too slowly to be in synchronous rotation.  We use a theory of tidal interaction in binary stars to place limits on the mass ratios that are independent of estimates based on the radial velocity curves.  The companions have masses below 0.26 $M_{\odot}$.  The pulsation spectra show the signature of high--overtone $g$-mode pulsation.  One star, KIC~11179657, has a clear sequence of $g$-modes with equal period spacings as well as several periodicities that depart from that trend.  KIC~02991403 shows a similar sequence, but has many more modes that do not fit the simple pattern.
%changed sentence 2 to clarify rotational splitting is the mechanism. Changed the to a in sentence 3.
\end{abstract}

\begin{keywords}
binaries: close -- stars: horizontal branch -- stars: rotation -- stars: oscillations -- subdwarfs
\end{keywords}

\section{Introduction}

Close binary systems are typically those in which the semi-major axis of the orbit is of comparable size to the radii of the stars themselves. In such systems, circularization of the orbit and synchronization of the components through tidal interactions is often assumed. The precise mechanisms of tidal coupling, and the resulting time-scales, have been under debate for a number of years.  Observational tests have been difficult to implement.  While orbital period determination is usually straightforward (through radial velocity curves and/or photometry), measuring the rotation of the component stars is challenging. Determinations of rotation periods using starspots becomes difficult in binary systems and impossible when those systems are evolved. It is possible to measure rotational broadening of spectral lines, but that requires metal lines which are inherently weak and need high S/N;  in binaries with compact components other broadening mechanisms swamp the rotational broadening signal for rotation rates of interest. When there are pulsations in one of the stars, asteroseismology can provide an alternative method for measuring rotation. The spinning of the star breaks degeneracy in the pulsation modes of non-radial pulsators causing single peaks to split into multiplets. This splitting can, through a simple relation, yield the rotation rate of the star \citep{Led51}.

One class of nonradially pulsating stars that can be ideal for this type of analysis is the pulsating subdwarf B(sdB) stars. These stars, remnants of low mass stars that have undergone helium core ignition,  lie along the extreme horizontal branch with temperatures 22000-40000 K and masses of $\approx$0.47$M_{\odot}$ \citep{saf94, heb84}. Their surface gravities ($\log g$ ranging from approximately 5.0 to 6.0) are typically higher than main sequence stars with corresponding temperatures. 
% This makes these stars difficult to observe with an average Kepler magnitude of 15.82 \citep{papI}. 
The formation process for a star with these characteristics is difficult to identify; one proposed formation channel is through common envelope ejection  \citep{han02,han03}. This implies that a significant number of these objects are members of binary systems. 

In addition, a selection of these stars pulsate nonradially. These pulsations fall into two distinct classes.  The  $p$-mode pulsators, also called V361 Hya stars, have periods that are generally in the 2-4 min range, with pulsation  amplitudes $\approx$ 1 per cent of their mean brightness \citep{kil07,reed07}. The $g$-mode pulsators, also known as V1093 Her stars, have pulsation periods between 45 min and 4 h and typical amplitudes less than 0.1 per cent of their mean brightness \citep{betsy03, royHelas10, reed11}.

Given their small pulsation amplitudes and multiperiodic nature, resolving the oscillation spectra of these $g$-mode pulsators is extremely challenging using ground--based facilities.  In particular, the periods of the $g$-mode pulsators result in only a few cycles for a single night's observing.  To use the pulsations to measure rotational effects requires identification of the azimuthal order of the modes, meaning that small frequency spacings between individual periodicities in the Fourier transform need to be measured unambiguously. Even with multi-site campaigns it is challenging to obtain a sufficient baseline for detailed study. 

% For this reason, although there is a theoretical basis for equal period spacing in g-modes \citep
% {Unno79} it was not well studied until \citep{reed11}. Even now, the theory is relatively 
% untested and needs extensive further study.

Recently, data on sdB pulsators provided by the \textit{Kepler} mission have spurred rapid progress in our understanding of these stars. The broader science goals, mission design, and overall performance of {\it Kepler} are reviewed by \citet{bor10} and \citet{koch10}; \citet{gillkap} review the asteroseismology component.  The {\it Kepler} Asteroseismic Science Consortium (KASC) working group on compact pulsators has made significant strides  toward our understanding of the  non-radial $g$-mode pulsators as a group \citep{papIII, reed11, gillkap, kaw10}, and have provided detailed seismic fits to two $g-$mode pulsators \citep{charp11, vang11}.  Many of these multiperiodic pulsators show distinct sequences of modes that are equally spaced in period, as is expected for high--overtone $g-$modes \citep{reed11}.

Two of the KASC targets are $g$-mode pulsators in close reflection--effect binary systems with periods of approximately 0.4 d: KIC 02991403 and KIC 11179657.  30~d of {\it Kepler} data from the survey phase,  hinted that neither system was in synchronous rotation, though the data durations were too short to fully resolve the pulsations \citep{kaw10}.  In a separate examination of an sdB star in the old open cluster NGC~6791, \citep[KIC 02438324, identified as ``B4'' by][]{kalud92}, \citet{pab11} used 6 months of {\it Kepler} data to show that it too is in a close reflection--effect binary.  B4 rotates with a period that is significantly longer than the orbital period of the system.  The similarity between these three systems suggests that the binary properties are closely related to their origin.

In this paper we report on results from longer duration observations of KIC 02991403 and KIC 11179657 with {\it Kepler}.  We identify several rotational triplets, demonstrating that the stars are not in synchronous rotation.   
%These results are consistent with the results of \citep{pab11} that $g$-mode pulsators in $\approx$ 0.5 day binary systems with low-mass M dwarf companions do not appear to be in synchronous rotation.
These results are consistent with the results of B4 shown in  \citep{pab11}, and suggests that 
% $g$-mode pulsators 
sdB stars in $\approx$ 0.5-d binary systems with low-mass M dwarf companions do not appear to be in synchronous rotation. 
KIC 11179657 shows a clear sequence of equally spaced periods as seen in many other $g$-mode sdB pulsators, including B4 \citep{pab11}. KIC 02991403 also shows a sequence of modes that share a constant spacing in period, but shows many more modes that do not follow this pattern.

\section{Observations}
%new sentence explainining the quarter system of Kepler

All observations were obtained by the \textit{Kepler} spacecraft between 2010 March and 2011 March. \textit{Kepler} data are made available in 3-~month spans and identified by quarter number. The observations we are analysing are from quarters Q5-Q8. The data were obtained in short cadence (SC) mode \citep{gillsc} which has a cadence of 58.85 s. Data were processed through the standard pipeline \citep{jonjendata}. All data products received through KASC have undergone Pre-search Data Conditioning. However, since this conditioning is optimized for exoplanet science, we used the simple aperture photometry flux. These fluxes are converted to fractional variations from the mean (i.e. differential intensity $\delta I/I$). The data show (mostly) subtle baseline changes from month to month, and so we treated each month of data individually, resulting in a light curve that is continuous across each quarter. We further edited the data by removing outliers; we clipped individual data points that departed from the (local) mean by 4$\sigma$. This resulted in a reduction of 1847 points (out of 486336) in KIC 02991403 and 947 points (out of 392532) in KIC 11179657.  
% we talked about this but i'm unsure what to add. Maybe something like, We have significant outliers making the number of points cut much higher than expected for the 4$/sigma$ level. 

Under normal operating circumstances, data received from \textit{Kepler} is continuous with the exception of small gaps, the largest of which is due to the quarterly roll of the spacecraft. For this reason {\it Kepler} has a duty cycle of over 90 per cent. 
% This in conjunction with randomly scattered outliers leads to a window function which is essentially a $sinc^{2}$. 
There are a small number of instrumental artefacts in the data commensurate with the long cadence (LC) readout rate (1/30 $\times$ SC) producing peaks at multiples of 566.44 $\mu$Hz, with the highest amplitude peaks generally lying in the 4000-7000 $\mu$Hz region \citep{gillsc}.  These artefacts are at a significantly higher frequency than the physical periodicities in these stars, and so do not impact our analysis.

We obtained 12 months of data (Q5-Q8) on KIC 02991403, resulting in a (formal) frequency resolution ($1/T$) of 0.032 $\mu$Hz.  The actual frequency precision achieved through least-squares fitting of sinusoids to the data \citep{monodon99} is significantly higher, but the higher precision is only realized if the individual periodicities are isolated from any other signal.  \citet{kaw10} identified 16 independent oscillation frequencies using 1 month of data from Q1. This star is a reflection effect binary with an orbital period of 10.633762 $\pm$ 0.000015 h.  
No significant signal is seen at the subharmonic of this periodicity, and the first harmonic is present at the 13\% level.  The phase of this harmonic peak is such that the maxima of the harmonic are at the extrema of the main peak, as is expected for a reflection-effect binary. 
Preliminary radial velocity measurements on the sdB star have an of amplitude 36 $\pm$ 2 km s$^{-1}$ \citep{tel11}. This velocity implies that the (unseen, except for the reflection effect) companion is lower in mass than the primary ($M < 0.47M_{\odot}$) if the inclination angle of the system is greater than 15$^{\circ}$. The absence of an eclipse places an upper limit on the inclination of approximately $80^{\circ}$, and places a lower limit on the companion mass of $\approx$ 0.089 $M_{\odot}$. The sdB star has $T_{\mathrm{eff}} =$ 27300 $\pm$ 200 K and $\log g = 5.43 \pm 0.03$ \citep{papI}.
%added a minimum mass for the companion determined by inclination constraints from the lack of an eclipse in this system. Don't understand the f(n) comment.
%changed less to greater...smaller the inclination the larger the companion mass.
% \subsection{KIC 11179657}

For the second target, KIC 11179657, we obtained 9 months of data (Q5-Q7); in Q8, KIC 11179657 fell on the beleaguered Module 3 in the {\it Kepler} focal plane which is no longer functional.  The resulting formal frequency resolution is 0.042 $\mu$Hz, though as noted above, isolated frequencies can be determined to much higher precision.  This star is also a reflection effect binary with an orbital period of 9.466936 $\pm$ 0.000024 h. It shows 8 oscillation frequencies in the 30-d survey data from Q2. The binarity in this reflection--effect system has also been confirmed via a measured velocity amplitude for the sdB star of 21 $\pm$ 2 km s$^{-1}$ \citep{tel11}. Assuming a non-zero inclination angle the companion in this system is again of low mass, with a minimum mass $\approx$ 0.047 $M_{\odot}$ if seen nearly edge--on; the companion exceeds the mass of the sdB if the inclination is less than approximately 8.5$^{\circ}$. The sdB star is slightly cooler and lower--gravity than its counterpart, with $T_{\mathrm{eff}}$ = 26000 $\pm$ 800 K,  and $\log g = 5.14 \pm 0.13 $ \citep{papI}.
%added a minimum mass and reworded slightly to address comment.

\section{Analysis}

We determined the frequencies of pulsation (and the orbital modulation period from the reflection effect) through Fourier analysis combined with non-linear least-squares fitting and prewhitening, as described in \citet{papII, papIII}; and \citet{kaw10}. This process identifies peaks in the Fourier transform and fits the amplitude, frequency and phase using a non-linear least-squares procedure. A sine curve with the fit parameters is then subtracted from the light curve to remove the periodicity. This process is then iterated until all significant peaks have been identified, fitted, and removed. A peak is defined as significant if it is more than 4$\sigma$ above the mean noise level. This was calculated by finding the mean value of the Fourier transform, outside of obvious peaks, in the $g$-mode region of each star.

Normal modes of oscillation in stars are characterized by the set of indices $n$, $l$ and $m$. The radial order $n$ provides a measure of the number of nodes in the radial direction for the given oscillation mode.  The angular indices $l$ and $m$ correspond to the spherical harmonic $Y_{l}^{m}(\theta,\phi)$.  For nonradial oscillations, in the absence of processes that break azimuthal symmetry, the oscillation frequencies depend only on $n$ and $l$; modes with different $m$ values are degenerate in frequency.  If the star is rotating (or has a sufficiently strong magnetic field), this can lift the $m$ degeneracy, revealing an equally spaced multiplet of $2l+1$ peaks if all $m$ components are present and if the rotation is sufficiently slow (or the magnetic field sufficiently weak).  For solid-body rotation with a period of $P_{\rm rot}$, the frequency spacing is proportional to the rotation rate:
\begin{equation}
  f_{n,l,m}=f_{n,l,0}+\frac{m}{P_{\rm rot}}(1-C_{n,l})
\end{equation}  
\citep{Led51} where $C_{n,l}$ is the Ledoux constant. $C_{n,l}$ is determined by the oscillation eigenfunctions. For models appropriate to KIC 02991403 and KIC 11179657 the range of $C_{n,l}$ is discussed in \cite{kaw10}, with $C_{n,1}$ between 0.465 and 0.496 while $C_{n,2}$ is between 0.158 and 0.165.  For these two targets, if the sdB components are in synchronous rotation, the rotational splitting should be approximately 13-15 $\mu$Hz for $l = 1$ modes.  

\subsection{KIC 02991403}

\begin{figure*}
%\plotone{gmodeft-02991403.eps}
\includegraphics{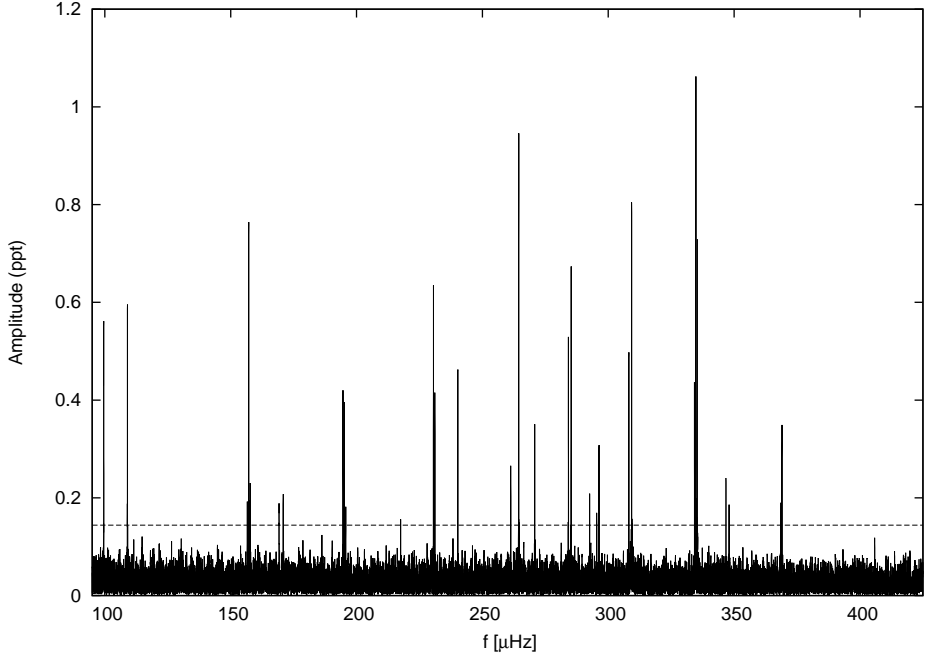}
\caption{The g-mode region of KIC 02991403. The 4-$\sigma$ significance threshold is represented by the dotted line. }
\label{FT-299}
\end{figure*}

With a full year of data, the number of detected oscillation modes has increased significantly compared to what was found by  \citet{papI}. All 16 modes previously identified are still present, and 18 additional periodicities lie above our significance threshold of 0.141 ppt (parts per thousand). The amplitude spectrum of the data is shown in Fig. \ref{FT-299}, and the significant periodicities are listed in Table \ref{freqtab299}.  In this table, frequencies identified in \citet{kaw10} are f1 through f16; the new periodicities, ordered by increasing frequency, are f18 through f34. The phase is given as an offset from $T_0$ of the first time of maximum (for that periodicity).
%changed wording of the last sentence. (5 on referee comments)

\begin{table*}
\caption{Periodicities of KIC 02991403.  Quoted errors are formal least-squares errors. Periodicities marked with an asterisk are slightly below the 4-$\sigma$ cutoff but included in the fit.}
\label{freqtab299}
\begin{tabular}{cccccll}
%\rotate
\hline
ID & Frequency  [$\mu$Hz] & Period [s] & Amplitude [ppt] & $T_{\rm max}^1$[s]& Orbital splitting & fine structure \\
\hline
\multicolumn{4}{c}{Binary period and first harmonic} \\
1f$_{\rm orb}$ & 26.122249(36) & 38281.542(51) & 17.657(35) &15971.0(24)\\
2f$_{\rm orb}$ & 52.24451(28) & 19140.8(10) & 2.279(35) &16062.0(93)\\
\\
f17 & 99.5625(11) & 10043.94(0.11) &0.562(35) &7122.0(200) & & \\
f18 & 108.994(11) & 9174.816(88) &0.601(35) &3352.0(170)& & \\
f19* & 114.8141(48) & 8709.73(0.37) &0.131(35) &5486.0(740)& & \\
f20 & 156.6102(41) & 6385.28(0.17) &0.153(35) &1678.0(460) & & \\
f1 & 157.17197(82) & 6362.458(34) &0.763(35) &2306.0(92) & &=f20+0.562 \\
f21 & 157.6653(35) & 6342.55(0.14) &0.195(36) &1549.0(380)& & \\
f22 & 157.7319(25) & 6339.87(0.11) &0.255(36) &5551.0(290) & &=f1+0.560 \\
f23 & 169.2203(32) & 5909.46(0.11) &0.200(35) &5618.0(330) & & \\
f24 & 170.8781(29) & 5852.124(99) &0.219(35) &3699.0(300) &=f22+13.146 & \\
f25* & 186.2481(48) & 5369.18(0.14) &0.128(35) &4935.0(460) & & \\
f2 & 194.6251(15) & 5138.083(39) &0.434(35) &2350.0(130) & & \\
f3 & 195.1789(15) & 5123.50(4) &0.417(35) &5055.0(140) & &=f2+0.554 \\
f26 & 195.7147(37) & 5109.477(97) &0.171(35) &1161.0(330) &=f24+24.837 &=f3+0.536 \\
f27 & 217.5162(41) & 4597.358(87) &0.154(35) &3131.0(330) & & \\
f4 & 230.554(10) & 4337.386(19) &0.626(35) &1671.0(77) &=f27+13.038 & \\
f5 & 231.1063(15) & 4327.014(29) &0.409(35) &2463.0(120) & &=f4+0.552 \\
f28 & 240.2369(14) & 4162.558(23) &0.466(35) &2530.0(99) & & \\
f29 & 261.2529(23) & 3827.709(34) &0.273(35) &479.0(150) & & \\
f6 & 264.43948(66) & 3781.5836(95) &0.956(35) &376.0(44) & & \\
f7 & 270.7766(18) & 3693.081(25) &0.345(35) &868.0(120) & & \\
f8 & 284.1718(12) & 3518.998(14) &0.541(35) &481.0(72) & & \\
f9 & 285.2697(93) & 3505.455(1) &0.684(35) &1768.0(57) & &=f8+1.098 \\
f30 & 292.5943(29) & 3417.702(34) &0.218(35) &3399.0(170) & & \\
f31 & 295.409(36) & 3385.137(42) &0.174(35) &2719.0(210) & & \\
f10 & 296.316(2) & 3374.78(23) &0.311(35) &3061.0(120) & &\\
f11 & 308.1784(12) & 3244.874(14) &0.484(35) &1164.0(74) & & \\
f12 & 309.27964(79) & 3233.3199(83) &0.800(35) &1470.0(45) & &=f11+1.101 \\
f13 & 334.2474(16) & 2991.796(14) &0.402(35) &2987.0(82) & & \\
f14 & 334.81955(59) & 2986.6835(51) &1.078(35) &728.0(31) & &=f13+0.572 \\
f15 & 335.37212(87) & 2981.7625(77) &0.729(35) &1943.0(45) & &=f14+0.553 \\
f32 & 346.7506(25) & 2883.917(21) &0.241(35) &71.0(130) & & \\
f33 & 347.995(33) & 2873.605(26) &0.194(35) &227.0(160) &=f14+13.145 &=f32+1.244 \\
f34 & 368.4757(36) & 2713.883(25) &0.178(35) &1188.0(170) & & \\
f16 & 369.0067(18) & 2709.978(12) &0.350(35) &1881.0(86) & & \\
\hline
\end{tabular}

\medskip
$^1$ $T_{0}$ is  BJD 2455276.4793502
\end{table*}

As noted in \citet{kaw10}, there is little evidence for frequency spacings between modes that match the expectations from synchronous rotation. The few splittings that are near the orbital frequency can be explained by the common period spacing in $l$= 1 $g$-modes (see below). In the range of frequencies typical for $g$-modes, the period spacing coincides with the system's orbital frequency.

\begin{figure*}
%\centering
\includegraphics{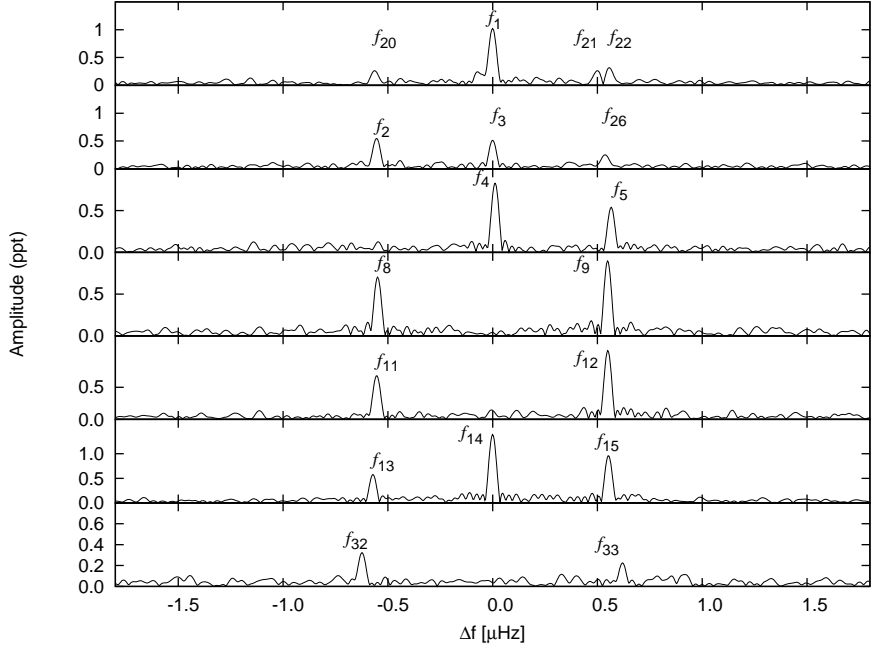}
%\plotone{peaks299.eps}
  \caption{Amplitude spectra of several multiplets in the $g$-mode region of KIC 02991403. The frequencies of each peak can be found with corresponding labels in Table \ref{freqtab299}. There are three triplets with an average spacing of 0.56 $\mu$Hz. Several doublets show twice this splitting.  There is some ambiguity in placement of f4 and f5 in the third panel from the top; they could be shifted to lower frequencies by 0.56 $\mu$Hz since we cannot uniquely assign $m$ values.} 
\label{stackedFT299}
\end{figure*}

With the improved frequency resolution and sensitivity available with 1 year of data, several equally-spaced triplets are now seen in the data, with spacing that is substantially smaller than the orbital frequency. One true triplet ($f$13, $f$14, $f$15) was identified by \citet{kaw10}, but  its spacing was near the run resolution. We now see several doublets and triplets with very similar spacings of $\approx$ 0.56 $\mu$Hz; these are shown in Fig. \ref{stackedFT299}. Of the 34 periodicities in Table \ref{freqtab299}, 16 are components of multiplets that show this spacing. The rotational period calculated using this frequency splitting assuming $l$ = 1 is $\approx$ 11 d. This is an order of magnitude longer than the orbital period, confirming that this system is not in synchronous rotation.  This ratio is very similar to that seen in B4 \citep{pab11}.  The possibility exists that the modes that are part of these multiplets with a 0.56 $\mu$Hz spacing are $l$ = 2 rather than $l$ = 1, even though no quintuplets are present.  If that were the case, the rotation period would be longer.
A rotation period of 11 d corresponds to a rotation velocity of approximately 1 km s$^{-1}$. This  rotation is slower than what is generally seen in  more ordinary (cooler) Horizontal Branch stars %with $T_{\rm{eff}}<$ 11500 
both in clusters and in the field \citep{behr03a, behr03b}. It is also consistent with spectroscopic measurements of many individual sdB pulsators \citep{heb00}. Since this is significantly longer than the orbital period, it is clear that not much spin up has occurred in these systems. 
%This is significantly slower than the mean rotation velocities of 5-7 km/s reported for sdB stars by \citet{geier11} suggesting that the sdB stars in binaries may be systematically slower rotators than sdB stars with no obvious companion.  In any case, it is clear that not much spin-up has occurred in these systems.

\begin{figure*}
%\centering
\includegraphics{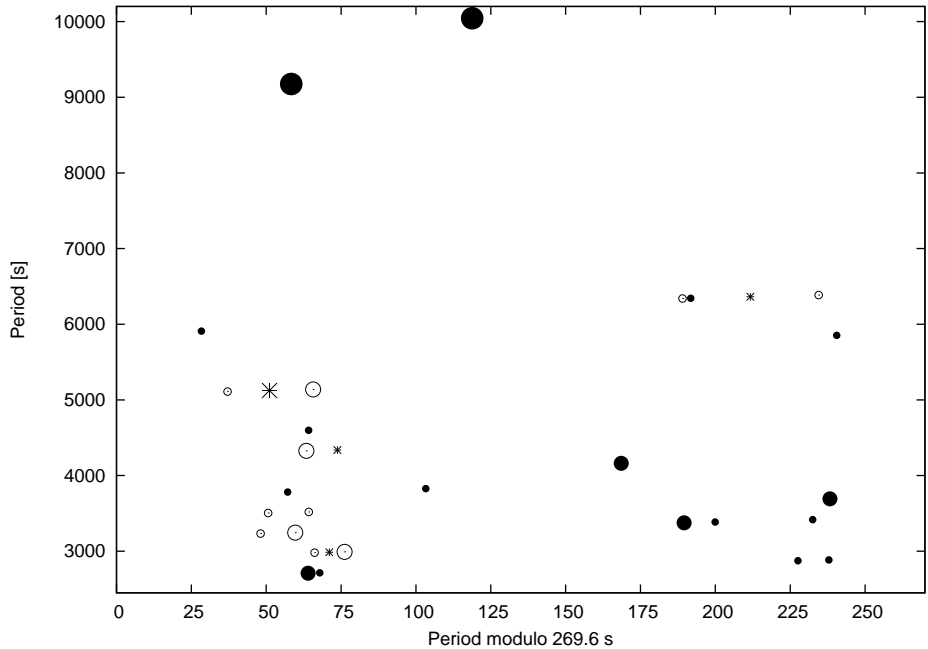}
%\plotone{299-echelle3.eps}
\caption{This {\'e}chelle diagram contains symbols for all periodicities identified in KIC~02991403.  The points are plotted modulo a fixed period spacing; those that have periods that are integer multiples of this period spacing will be aligned vertically.  
Stars denote $m = 0$ modes (those at the centre of triplets), and open circles are suspected $m = \pm 1$ modes.  Filled circles are isolated periods. Larger symbols indicate higher amplitude peaks.  A vertical column centered near 50-75 s contains most of the identified triplets. There are also several points that  lie far away from this grouping, and may be indicative of higher $l$ modes, or modes that are strongly affected by internal composition gradients. There is an artificial offset of 50 seconds imposed in the figure so the points do not fold across 0 s. }
\label{ech-299}
\end{figure*}

In many $g-$mode sdB pulsators, the periods of the modes display period spacings that are integer multiples of a common period spacing \citep{reed11}, as expected from asymptotic theory for models without internal compositional discontinuities.  To examine the possibility that KIC 02991403 shows this signature, we attempted to find a common period spacing. For $l$ = 1 modes in sdB stars, this spacing has been observationally determined to be remarkably uniform; most stars show a spacing near 250 s, which is consistent with the asymptotic $l$ = 1 spacing from theoretical models \citep{reed11}.  \citet{kaw10} explored this question with one month of data and concluded that there might be a mix of $l$ = 1 and $l$ = 2 modes in KIC 02991403.  With these new data, most of the periods corresponding to the centres of the triplets in Fig. \ref{stackedFT299} seem consistent with a spacing of 269.6 s.  Folding all identified periodicities at that spacing leads to the {\'e}chelle diagram shown in Fig. \ref{ech-299}. A likely $l$ = 1 sequence seems to be present near 60 s in this figure; however, many peaks are not associated with this group.  Theoretical models of high--overtone $g-$modes \citep[see, for example][]{charp11, vang11} show periods that approximate this behavior, though  large departures (up to 100 s) are common.  These departures from uniform period spacing are the consequence of internal composition transition zones, which lead to mode trapping and departure from the asymptotic relationship.  In KIC~02991403, the spread of points in the {\'e}chelle diagram (Fig. \ref{ech-299}) may be because some of the periodicities are $l$ = 2 modes, or that the internal structure of this star contains features that cause mode trapping.  Thus the problem remains as far as clearly identifying $l$ for many of the modes of KIC 02991403.

\subsection{KIC 11179657}

\begin{figure*}
%\plotone{gmodeft-11179657.eps}
\includegraphics{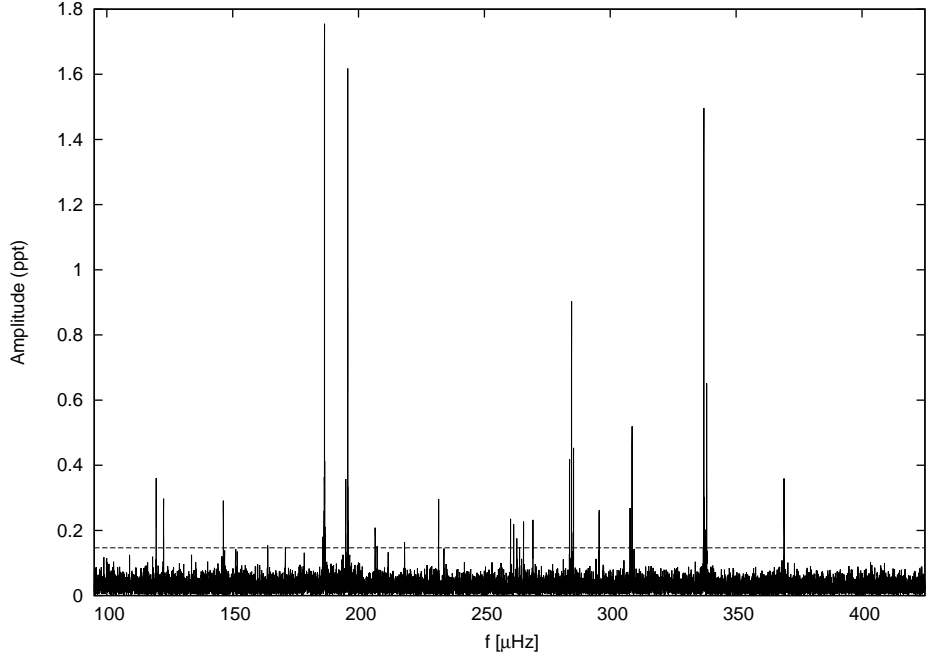}
 \caption{The $g-$mode region of KIC 11179657. The 4-$\sigma$ level above the noise is represented by the dotted line.}
\label{FT-111}
\end{figure*}

Nine months of data {\it Kepler} photometry on KIC~11179657 brings the 4-$\sigma$ noise threshold down to an amplitude of 0.146 ppt.  At this level, combined with the frequency resolution gain, we can identify 18 periodicities in the light curve in addition to the 8 periodicities found by \citet{papI}; two that were close to the detection threshold in \citet{kaw10} are confirmed. Fig. \ref{FT-111} shows the amplitude spectrum of the new data over the range of frequencies where significant periodicities are found.  Table \ref{freqtab111} presents the frequency list; we use the same naming convention as used for KIC 02991403.

\begin{table*}
\caption{Periodicites of KIC 11179657.  Quoted errors are formal least-squares errors. Periodicities marked with an asterisk are slightly below the 4-$\sigma$ cutoff but included in the fit.}
\label{freqtab111}
\begin{tabular}{cccccll}
%\rotate
\hline
ID & Frequency  [$\mu$Hz] & Period [s] & Amplitude [ppt] & $T_{\rm max}^1$[s]& Orbital splitting & fine structure \\
\hline
\multicolumn{4}{c}{Binary period and first harmonic} \\
\\
1f$_{\rm orb}$ & 29.341887(74) & 34080.971(87) & 9.36(3) &19462.0(35.0)\\
2f$_{\rm orb}$ & 58.6836(11) & 17040.53(0.31) & 0.66(3) &2582.0(250.0)\\
\\
f9 & 119.6149(19) & 8360.16(0.13) &0.36(0.03) &1152.0(220) & & \\
f10 & 122.5558(24) & 8159.55(0.16) &0.29(0.03) &838.0(270) & & \\
f11 & 146.2958(24) & 6835.47(0.11) &0.29(0.03) &560.0(220) & & \\
f12 & 185.7243(35) & 5384.3(0.10) &0.20(0.03) &5382.0(260) & & \\
f1 & 186.4778(39) & 5362.569(1) &1.77(0.03) &5157.0(29) & &=f12+0.753 \\
f13 & 194.925(2) & 5130.169(51) &0.35(0.03) &5002.0(140) & & \\
f2 & 195.72284(43) & 5109.266(1) &1.63(0.03) &3465.0(30) & &=f13+0.798 \\
f14 & 206.5921(33) & 4840.455(78) &0.21(0.03) &139.0(220) & & \\
f15 & 218.2891(41) & 4581.081(87) &0.17(0.03) &2368.0(260) & & \\
f16 & 231.8228(23) & 4313.64(0.43) &0.30(0.03) &1005.0(140) & & \\
f17 & 260.39(3) & 3840.391(45) &0.23(0.03) &2825.0(160) & & \\
f18 & 261.6353(32) & 3822.114(47) &0.22(0.03) &185.0(170) &=f16+29.8125 &=f17+1.245 \\
f19 & 262.873(4) & 3804.117(58) &0.17(0.03) &2051.0(210) & &=f18+1.238 \\
f20 & 265.5517(33) & 3765.746(46) &0.21(0.03) &861.0(170) & &=f19+2.679 \\
f21 & 269.278(3) & 3713.634(42) &0.23(0.03) &2098.0(160) & & \\
f22 & 283.83(17) & 3523.235(21) &0.42(0.03) &2681.0(81) & & \\
f3 & 284.62843(78) & 3513.3525(96) &0.90(0.03) &3283.0(37) & &=f22+0.798 \\
f4 & 285.4093(16) & 3503.74(0.02) &0.44(0.03) &591.0(77) & &=f3+0.781 \\
f5 & 295.5828(25) & 3383.15(0.03) &0.27(0.03) &890.0(120) & & \\
f23 & 307.8559(25) & 3248.273(28) &0.27(0.03) &1018.0(120) & & \\
f6 & 308.6558(14) & 3239.855(14) &0.52(0.03) &1356.0(60) & &=f23+0.800 \\
f24$^*$ & 309.4427(48) & 3231.62(0.05) &0.14(0.03) &71.0(210) & &=f6+0.787 \\
f7 & 337.17368(46) & 2965.8306(41) &1.48(0.03) &1963.0(19)  &=f23+29.3178 & \\
f25 & 337.9506(38) & 2959.013(33) &0.19(0.03) &470.0(150) &=f6+29.2948 &=f7+0.777 \\
f8 & 338.2984(12) & 2955.97(0.10) &0.60(0.03) &698.0(47) &=f6+29.6426 &=f7+1.125 \\
f26 & 369.0297(19) & 2709.809(14) &0.36(0.03) &2115.0(73) & & \\
\hline
\end{tabular}

\medskip
$^1$ $T_{0}$ is  BJD 2455276.4801154
\end{table*}

\begin{figure*}
%\centering
\includegraphics{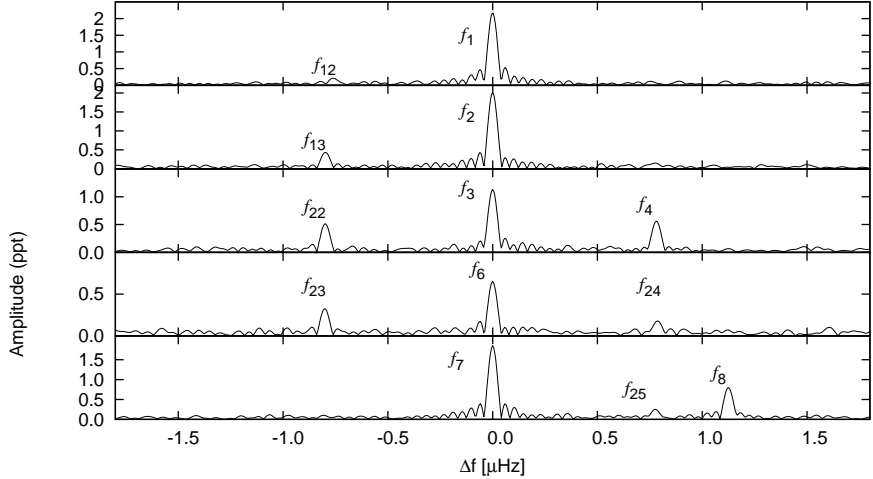}
%\plotone{peaks111.eps}
  \caption{Rotationally split multiplets in KIC~11179657 associated with the suspected $l$ = 1 spacing $\approx$  0.784 $\mu$Hz. The frequencies of each peak can be found with corresponding labels in Table \ref{freqtab111}. There are two triplets (middle and second to bottom) with an average spacing of 0.793 $\mu$Hz, though we note that f24 is slightly below our formal detection threshold. There are several doublets which show this splitting. In the bottom panel centered on f7, f25 is spaced at the $l$ = 1 rotational splitting distance. The f8 peak is separated by 1.12$\mu$Hz from f7, which is smaller than the expected $l$ = 2 splitting.}  
\label{stackedFT111}
\end{figure*}

\citet{kaw10} failed to identify any equally-spaced triplets using the one-month survey data. With the extended data from Q5-Q7, we find that all but one of the previously detected frequencies (f5) is a member of a triplet or pairs with another frequency at twice the triplet splitting.  These common frequency spacings are much smaller than the orbital frequency. The most common spacing is approximately 0.78 $\mu$Hz; 12 periodicities are parts of multiplets with this splitting. The uniformity of this splitting is shown in Fig. \ref{stackedFT111} where we expand the Fourier transform around each multiplet.  In two instances, the central peak has the largest amplitude, and we align the incomplete multiplets assuming the large amplitude component is the central one. The labelled peaks are all above or near our significance threshold as indicated by the number corresponding to the entry in Table \ref{freqtab111}.  These modes are split by the same spacing of 0.78 $\mu$Hz. 

Four other periodicities (f17, f18, f19, and f20) show a spacing of $\approx$ 1.29 $\mu$Hz. The ratio between the 0.78 $\mu$Hz and 1.29$\mu$Hz splitting is 1.65, consistent with that expected for $l$ = 1 modes (0.78 $\mu$Hz) and $l$ = 2 modes (1.29 $\mu$Hz).  Thus the four frequencies appear to be part of an $l$ = 2 quintuplet which is missing one peak. The calculation of the rotational period from  these spacings is consistent with  $\approx$ 7.4 d which, as in KIC 02991403 and B4, is much longer than the orbital period.  We note that the frequency spacing between f7 and f8 is close to, but significantly smaller than, the $l$ = 2 spacing.

%This is further evidence that that sdB stars which share this parameter space should not be in synchronous rotation.

\begin{figure*}
%\centering%
\includegraphics{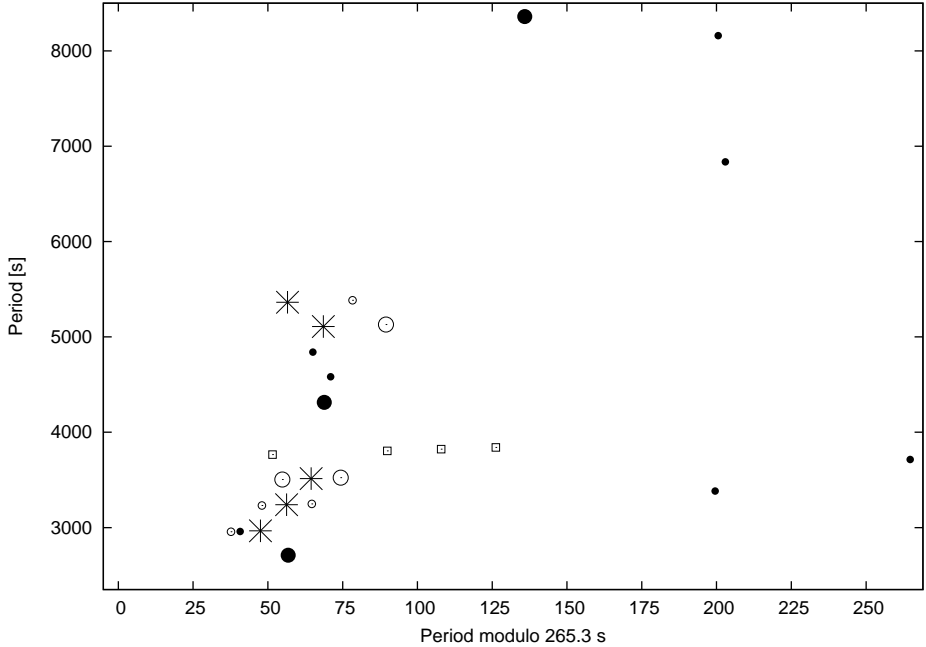}
%\plotone{111-echelle3.eps}
  \caption{{\'E}chelle diagram of the periodicities $f$1-$f$26 of KIC 11179657 with a folding period is 265.3 s. Filled circles are undetermined peaks, stars are $l$ = 1,  $m$ = 0 modes, open circles $l$ = 1 $m=\pm$ 1 modes, and open squares are suspected members of an $l= 2$ multiplet. The symbol size is proportional to peak amplitude. There is a clear grouping around 70 s which is most likely associated with $l$ = 1. The members of the  $l$ = 2 multiplet are offset from the $l$= 1 ridge, though the highest frequency member (and the missing adjacent mode) do overlap with the $l$ = 1 ridge. This overlap is not unexpected as they can have different values of $n$.}
\label{ech-111}
\end{figure*}

Unlike KIC 02991403, most of the periodicities of KIC 11179657 have a consistent period spacing. The period spacing found in this star is 265.3 s and appears to be fairly uniform. In Fig. \ref{ech-111} we see a well defined column of periodicities that are all consistent with being $l$ = 1 modes. A few of the periodicities are offset from the ridge by about 140 seconds, which may indicate modes that are affected by mode trapping by internal composition transition zones (see above).  Note that in Fig. \ref{ech-111} the suspected $l$ = 2 multiplet is also offset from the $l$ = 1 ridge.
%, and that f7 is on the $l$=1 ridge. I can't figure out why this is here either so i'm erasing it.  At this stage, we await detailed seismic modeling of these stars to understand the features of this rather clean echelle diagram.

%This is the paragraph where I most need your input about the mode-trapping

\section{Synchronization}

Synchronization has not been reached in either KIC 02991403 or KIC 11179657, meaning that the synchronization time-scale must be significantly longer than the time that these systems have been in their current configuration.  The time-scale for life as a helium core burning sdB star is $10^8$ years.  Without more precise models of these stars (constrained by seismology) it is difficult to know their ages, but we can estimate the synchronization times using two prescriptions. KIC 02991403 and KIC 11179657 have very similar system parameters in terms of masses, rotation rates, and orbital periods.  Therefore, the analysis below applies to both systems.

The model of  \citet{Tas97} produces a synchronization time that is extremely short for sdB+dM binaries with these periods (as discussed in \citealt{pab11}). On the other hand, synchronization through coupling via dynamical tides, as proposed by \citet{Zahn75}, produces an approximate parametrized synchronization time-scale \citep{clcu97, pab11}.
\begin{equation}
\tau_{\rm syn}=3.43 \times 10^{6}\mbox{yr}\left(\frac{\beta}{0.13}\right)^{2}\left(\frac{1+q}{q}\right)^{2}\left(\frac{M}{M_{\odot}}\right)^{\frac{7}{3}}
\left(\frac{R}{R_{\odot}}\right)^{-7}\left(\frac{P}{{\rm day}}\right)^{\frac{17}{3}}\left(\frac{E_{2}}{10^{-8}}\right)^{-1}
\end{equation}  
where $\beta$ is ``the radius of gyration" (the moment of inertia divided by $MR^{2}$), $q$ is the mass ratio,  and $E_{2}$ is the tidal constant, which is highly dependent on stellar structure. $E_{2}$ is difficult to calculate directly, but is proportional to  $\left( r_{c} / R\right)^{8}$  where $(r_c/R)$ is the fractional radius of the convective core \citep{Zahn77}. Scaling from main sequence models by Claret \citep{clcu97} we find that, to within a factor of three, $E_{2}$ = $\left( r_{c} / R\right)^{8}$ for sdB stars. The error introduced is small compared with our other uncertainties. As the sdB star evolves, $(r_c/R)$, $R$, and (to a lesser extent) $\beta$ can change causing $\tau_{syn}$ to vary.  

This approach to synchronization can be integrated using the computed $\tau_{\rm syn}$ as a function of time via 
\begin{equation}
\frac{1}{\tau_{\rm syn}}=\frac{1}{\Omega-\omega}\frac{d\omega}{dt}
\end{equation}
where $\Omega$ is the orbital angular velocity and $\omega$ is the star's angular rotation velocity. %(note that definition of $\Omega$ differs from its earlier use in this paper). There should be a reference in the Ledoux equation but it has been changed.
Using the value of $\tau_{\rm syn}$ as a function of age allows us to calculate $\omega(t)$. When $\omega$ $\approx$ $\Omega$ this system is synchronized. Conservation of total angular momentum implies that spinning up the sdB star will come at the expense of orbital angular momentum. However, since the rotational inertia of the star is less than 1 percent of the orbital angular momentum, spin-up of the star will occur on a time scale that is approximately 100 times shorter than the rate of change of the orbital angular momentum, regardless of the companions mass.  We therefore assume that $\Omega$ is constant.  We also assume that the secondary star is in synchronous rotation so its angular momentum evolution can be ignored. We can also compute the synchronization time as a function of the mass ratio $q$ for comparison with the data -- smaller mass ratios imply longer synchronization times. With a typical mass for the sdB star of $\approx$ 0.47 $M_{\odot}$, and values for $R$ and $(r_c/R)$ from representative sequences of evolutionary models \citep{2010c,kahos}, the fact that these stars are not yet synchronized (at an upper limit to the age of $10^8$ years)  we should be able to place limits on the mass of the companion.For values of $E_{2}$ greater than $\left( r_{c} / R\right)^{8}$, if the age of the system is half the mean lifetime ($5\times 10^7$ y) then $M_{c} < 0.26 M_{\odot}$. On the other hand, if $E_{2}$ is much less than $\left( r_{c} / R\right)^{8}$ then it is likely that the system will never spin-up. In that case  for the system to have spun-\textit{up} to the value observed starting from a rotation period of 100 d,  $M_{c}$ must be greater than 0.16 $M_{\odot}$. The value of $M_{c}$ is valid for rotation periods longer than about 50 d.

\section{Discussion}
 
KIC 02991403 and KIC 11179657 are two very similar sdB $g$-mode pulsators in binary systems. Both systems have several triplet spacings, which imply rotation periods of  10.3 d and 7.4 d, respectively. This assumes solid body rotation, but as the $g$-modes are sensitive to virtually all depths in the star, our conclusions should change little. Since the orbital period in both systems is $\approx$ 0.4 d both systems are in non-synchronous rotation. These  properties are shared with the sdB binary in NGC~6791 investigated by \citet{pab11}.  These results can place tight constraints on theoretical models for tidal synchronization.  For example, the \citet{Tas97} mechanism would synchronize these stars in a very short time.  

On the other hand, if the analysis following \citet{Zahn77} and \citet{pab11} in the previous section is accurate, then we can use the {\it lack} of synchronization to place limits on the mass ratio and therefore on the mass of the secondary star. This mass limit, along with the radial velocity amplitude, can in turn provide a lower limit on the orbital inclination. If $E_{2} > \left( r_{c} / R\right)^{8}$ this turns out to be $\approx 23^{\circ}$ for KIC 02991403 and $\approx 13^{\circ}$ for KIC 11179657. However, if $E_{2} < \left( r_{c} / R\right)^{8}$ then we can set a maximum inclination of $\approx 35^{\circ}$  for KIC 02991403 and $\approx 19^{\circ}$ for KIC 11179657. Using the fact that there are no eclipses we can set a independent maximum inclination that is $\approx 80^{\circ}$ for both stars. While the estimates are rough the synchronization and spectroscopy yield consistent results. 

%Added extra info comparing synchronization and spectroscopy

%Using this we can place an upper limit on the mass of the companion of $M_{c}$ $<$ 0.22 $M_{\odot}$ confirming the spectroscopic results of (Ostenson and Telting REFERENCE). 

Intriguingly, despite their similar parameters, the period distributions in these stars are quite different. The periods are well-behaved, with a clear period spacing, $l$ = 1 multiplets and, in the case of KIC 11179657, an $l$ = 2 multiplet. While KIC~2991403 does show an $l = 1$ sequence, there is no clear evidence of $l$ = 2 multiplets despite having a longer baseline and more pulsation modes than KIC 11179657.  Compared to KIC 11179657, KIC 02991403 shows a large number of periodicities that are clearly separated from the $l$ = 1 ridge, suggesting that it has several $l$ = 2 modes that are not parts of multiplets and/or it has a significant number of modes that are influenced by internal composition transition zones.

We expect to obtain at least two years of data on these targets, and perhaps much more, over the course of the {\it Kepler} mission. With more data, lower-amplitude modes may be revealed that could complete additional multiplets.   
% However, this will not obviously fix the weak period spacing that we expect in this system. 
Given the current observed modes, however, seismic modelling may be able to reveal subsurface composition transitions that would explain the $g$-mode period distribution.

%disturbed is another comment I'll leave up to you to answer. 

\section*{Acknowledgments}
Funding for this Discovery mission is provided by NASA's Science
Mission Directorate. The authors gratefully acknowledge the entire {\sl Kepler} team, whose efforts
have made these results possible. The authors also acknowledge the KITP staff of UCSB for their warm hospitality during the research program ``Asteroseismology in the Space Age.'' This KITP program  was supported in part by the National Science Foundation of the United States under Grant No.\ NSF PHY05--51164. The research leading to these results has also received funding from the
European Research Council under the European Community's Seventh Framework Programme (FP7/2007--2013)/ERC grant agreement n$^\circ$227224 (PROSPERITY), as
well as from the Research Council of K.U.Leuven grant agreement GOA/2008/04. Haili Hu is supported by the Netherlands Organisation
for Scientific Research (NWO). Steven Bloemen acknowledges the travel grant (n$^\circ$V446211N) he received from Fund for Scientific Research of Flanders (FWO) for his stay at KITP.

%added the sentence about KITP using Connie's suggestion. 

%\begin{thebibliography}{12}

\label{lastpage}

\end{document}